\documentclass[
 reprint,
 amsmath,amssymb,
 aps,
 prd,
]{revtex4-1}

\usepackage{xcolor}
\usepackage{graphicx}
\usepackage{dcolumn}
\usepackage{bm}
\usepackage{hyperref}

\def\bea{\begin{eqnarray}}
\def\eea{\end{eqnarray}}

\date{\today} 
\begin{document} 

\title{The de Sitter Instanton from Euclidean Dynamical Triangulations}

\author{Scott Bassler}
\affiliation{Department of Physics, Syracuse University, Syracuse, NY 13244}
\author{Jack Laiho}
\email{jwlaiho@syr.edu}
\affiliation{Department of Physics, Syracuse University, Syracuse, NY 13244}
\author{Marc Schiffer}
\email{schiffer@thphys.uni-heidelerg.de}
\affiliation{Institut f\"{u}r Theoretische Physik, Universit\"{a}t Heidelberg,
Philosophenweg 16, 69120 Heidelberg, Germany}
\author{Judah Unmuth-Yockey}
\affiliation{Department of Physics, Syracuse University, Syracuse, NY 13244}
\affiliation{Fermi National Accelerator Laboratory, Batavia, IL 60510}

\begin{abstract}

We study the emergence of de Sitter space in Euclidean dynamical triangulations (EDT).  Working within the semi-classical approximation, it is possible to relate the lattice parameters entering the simulations to the partition function of Euclidean quantum gravity.  We verify that the EDT geometries behave semi-classically, and by making contact with the Hawking-Moss instanton solution for the Euclidean partition function, we show how to extract a value of the renormalized Newton's constant from the simulations.  This value is consistent with that of our previous determination coming from the interaction of scalar particles.  That the same universal constant appears in these two different sectors of the theory is a strong indication that EDT provides a viable formulation of quantum gravity.

\end{abstract}

\maketitle

\section{Introduction}
\label{sec:intro}

The quantization of gravity remains one of the great outstanding problems in theoretical physics.  In this work we continue our studies of Euclidean dynamical triangulations (EDT) \cite{Ambjorn:1991pq, Agishtein:1991cv, Catterall:1994pg}, a lattice approach to formulating quantum gravity.  This approach attempts to make contact with the asymptotic safety scenario of Weinberg \cite{Weinberg:1980gg}, where the existence of a nontrivial ultraviolet fixed point would make the theory effectively renormalizable nonperturbatively.  The perturbative nonrenormalizability of quantum gravity is well known \cite{tHooft:1974toh, Goroff:1985th}.  In order to realize the asymptotic safety scenario, EDT would have to recover classical general relativity in the appropriate limit, and there would have to exist a continuous phase transition in the phase diagram of the lattice theory such that a divergent correlation length would allow one to take the continuum limit.  

We briefly review here the evidence that EDT satisfies these conditions.  Ref.~\cite{Laiho:2016nlp} showed that a fine tuning of the exponent of a local measure term is needed in order to recover physical results.  This local measure term was first introduced into the lattice theory some time ago in Ref.~\cite{Bruegmann:1992jk}, but the evidence that this term is needed to recover semiclassical physics and to take the continuum limit has been presented more recently\cite{Laiho:2016nlp}.  Once the tuning procedure is implemented, it is found that the geometries in the simulations are consistent with being four-dimensional, and their behavior is close to that of Euclidean de Sitter space.  There is a well-established first order phase transition in the phase diagram \cite{Bialas:1996wu, deBakker:1996zx, Ambjorn:2013eha, Coumbe:2014nea, Rindlisbacher:2015ewa}.  Adding the measure term introduces a new parameter, thus enlarging the phase diagram and turning the first order point into a first order line. There appears to be no obstacle to taking the continuum limit by following this first order line to a possible critical endpoint; ensembles following this procedure were generated at a number of different lattice spacings \cite{Laiho:2016nlp, Dai:2021fqb}.  

The evidence for semiclassical physics seen in Ref~\cite{Laiho:2016nlp} is the following.  The global Hausdorff dimension was measured using finite-volume scaling, and it was shown to be close to four \cite{Laiho:2016nlp}.  The spectral dimension, a fractal dimension defined by a diffusion process, varies with distance scale, and it was shown to approach a value close to four at long distances.  This variation of the spectral dimension with distance had been seen already in other approaches \cite{Ambjorn:2005db, Lauscher:2005qz, Carlip:2017eud}.  It was also shown in Ref.~\cite{Laiho:2016nlp} that the average over geometries gives a result that is close to that of Euclidean de Sitter space, and the quantitative agreement with the classical solution gets better as the proposed continuum limit is approached.  The agreement between the classical solution and the lattice data is actually the worst at long distances, but it improves as the lattice spacing is reduced.  This might seem counter-intuitive, since it is typically the short-distance behavior that is modified by discretization effects, but this type of effect on long-distance behavior is common when a symmetry of a theory is broken by the regulator, for example by the finite lattice spacing in the case of lattice regularization.  An example of this is the Wilson fermion formulation of lattice quantum chromodynamics (QCD), where the lattice regulator breaks chiral symmetry.  There a fine-tuning is required to restore the symmetry, and even then, at finite lattice spacing the chiral symmetry breaking leads to distortions of the pion sector, which contains the lightest (and therefore longest Compton wavelength) states of the theory.  Ref.~\cite{Laiho:2016nlp} argued that the analogous symmetry that is broken by dynamical triangulations is continuum diffeomorphism invariance. 

 Calculations including matter fields (in the quenched approximation) also provide evidence for the emergence of semiclassical spacetimes with the hoped-for behavior.  Ref.~\cite{Catterall:2018dns} introduced K\"{a}hler-Dirac fermions \cite{Kahler:1962} to the EDT formulation.  This approach provides a generalization of staggered fermions to the random lattices of dynamical triangulations without the need to introduce vielbeins or spin connections.  In the flat-space, continuum theory the K\"{a}hler-Dirac action reduces to four copies of Dirac fermions \cite{Banks1982,Catterall:2018lkj}.  This appears to be true of K\"{a}hler-Dirac fermions coupled to EDT as well, but only in the continuum, infinite-volume limit.  This is seen in the approximate four-fold degeneracy of the low-lying eigenvalues of the K\"{a}hler-Dirac matrix, and in the degeneracy of scalar bound states in the continuum limit \cite{Catterall:2018dns}.  The four-fold degeneracy is lifted by lattice discretization effects in a similar manner to what is found with staggered fermions in lattice QCD \cite{Bazavov:2009bb}, but just as in lattice QCD, the degeneracy appears to be restored in the continuum limit.  An additional advantage of K\"{a}hler-Dirac fermions is that they possess an exact $U(1)$ symmetry, which is related to continuum chiral symmetry.  A study of fermion bilinear condensates provides strong evidence that this $U(1)$ symmetry is not spontaneously broken at order the Planck scale, implying that fermion bound states do not acquire unacceptably large masses due to chiral symmetry breaking.  These results for K\"{a}hler-Dirac fermions in EDT show that lattice fermions with the desired properties can be incorporated into the theory.

 Ref.~\cite{Dai:2021fqb} followed up this work with a study of the gravitational interaction of scalar fields.  Once again working in the quenched approximation, Ref.~\cite{Dai:2021fqb} studied the binding energy of two scalar particles in the non-relativistic limit, as originally proposed in Ref.~\cite{deBakker:1996qf}.  By looking at the binding energy as a function of the constituent particle mass, it was shown in Ref.~\cite{Dai:2021fqb} that the ground-state energy of the bound-state system is compatible with the result of solving the Schr\"{o}dinger equation with Newton's potential in 3+1 dimensions.  This recovery of Newton's law of gravitation in the appropriate limit allowed the determination of the renormalized Newton's constant $G$ for the first time within EDT.  This value of $G$ sets the lattice spacing in units of the Planck length, and given the value that was obtained, it was determined that the lattice spacings of the simulations are smaller than the Planck length.  This suggests that there is no barrier to taking the continuum limit.

 Given these successes in the quenched matter sector, it is interesting to return to a study of the global behavior of the geometries.  As noted above, the geometries resemble Euclidean de Sitter space.  This is seen by comparing the classical de Sitter solution (analytically continued to imaginary time) to the ensemble average of the shape of the geometries.  The agreement between the lattice data and the expected classical, continuum curve gets better as the continuum limit of the lattice theory is approached \cite{Laiho:2016nlp}.  Even so, it would be desirable to have additional cross-checks that the lattice geometries are actually approaching semiclassical de Sitter space, and that the simulations properly account for quantum fluctuations about the classical solution.  In this regard we take our inspiration from causal dynamical triangulations (CDT), a variant of the dynamical triangulations approach in which a foliation of the geometries is introduced explicitly \cite{Ambjorn:1998xu, Ambjorn:2001cv, Loll:2019rdj}.  

 Many of the nice properties that appear to be recovered by EDT in the continuum limit were first seen in CDT, including the emergence of four-dimensional (Euclidean) de Sitter space \cite{Ambjorn:2004qm, Ambjorn:2005qt, Klitgaard:2020rtv} and the scale-dependence of the spectral dimension \cite{Ambjorn:2005db}.  In the CDT formulation it has also been shown that semiclassical fluctuations about de Sitter space are well-described by a simple mini-superspace model \cite{Ambjorn:2007jv, Ambjorn:2008wc}.  Thus, the evidence for the emergence of four-dimensional de Sitter space in CDT is quite compelling.  The EDT geometries in our simulations have baby universe-like structures that have a cross-section of order the cut-off but are rather long in linear extent.  These baby universes branch off of the mother universe, and they seem to cause a large deviation from the putative de Sitter solution at finite lattice spacing.  Nonetheless, this deviation gets smaller and appears to vanish as the EDT continuum limit is approached \cite{Laiho:2016nlp}, though the study of quantum fluctuations about de Sitter space is complicated by these effects at finite lattice spacing.  Since branching baby universes are explicitly forbidden in the CDT path integral, that formulation does not suffer from this particular problem.

 In this work we look at the finite-volume scaling of one of the bare parameters in the lattice action, the bare cosmological constant.  It can be shown that if semiclassical physics is to be recovered, then this bare parameter should be a linear function of $1/\sqrt{V}$, where $V$ is the lattice volume.  We show that for large volumes our lattice data is in fact consistent with this expectation.  Following the discussion in Ref.~\cite{Ambjorn:2012jv}, we study the saddle-point approximation of the Euclidean partition function about de Sitter space, and we show how the parameters in the effective action for the lattice theory can be related to the continuum Hawking-Moss instanton solution \cite{Hawking:1981fz}, evaluated for the special case of de Sitter space.  This relationship allows us to use the finite-size scaling of our bare cosmological constant to obtain a result for the renormalized Newton's constant.  Our result is in excellent agreement with the previous determination of Newton's constant from the interaction of scalar particles in Ref.~\cite{Dai:2021fqb}, providing further evidence that the EDT formulation realizes a theory of gravity, and that de Sitter space with the correct quantum fluctuations emerges from our simulations.  The determination of $G$ in this work is to a higher precision than that of Ref.~\cite{Dai:2021fqb}, allowing for a slightly better determination of our absolute lattice spacing.

 This paper is organized as follows:  Section \ref{sec:edt} reviews the EDT formulation.  Section \ref{sec:desitter} discusses the de Sitter instanton solution and how it may be used to extract the renormalized Newton's constant in our framework.  Section \ref{sec:sim} gives the details of the simulations.  Section \ref{sec:num} presents our numerical results for the finite-size scaling of the bare cosmological constant and our determination of $G$, as well as our results for the lattice distance conversion factors needed to complete our calculation of $G$ and to compare it to previous results.  We conclude in Section \ref{sec:conclusion}.

\section{Euclidean Dynamical Triangulations} 
\label{sec:edt}

In Euclidean quantum gravity the partition function is given by the path-integral sum over geometries,
\bea\label{eq:part} Z_E= \int {\cal D}[g] e^{-S_{EH}[g]},
\eea
where the Euclidean Einstein-Hilbert action is
\bea\label{eq:ERcont} S_{EH}=-\frac{1}{16\pi G}\int d^4x\sqrt{g}(R -2\Lambda),
\eea
with $R$ the curvature scalar, $\Lambda$ the cosmological constant, and $G$ Newton's constant.

The EDT approach to quantum gravity assumes that the partition function is given by the sum over triangulations \cite{Ambjorn:1991pq, Bilke:1998vj}
\bea\label{eq:Z} Z_E = \sum_T \frac{1}{C_T}\left[\prod_{j=1}^{N_2}{\cal O}(t_j)^\beta\right]e^{-S_{ER}},
\eea
where the factor $C_T$ divides out equivalent ways of labeling the vertices in a given geometry, the term in brackets is a local measure term with the product over all triangles, and ${\cal O}(t_j)$ is the order of triangle $j$, i.e. the number of four simplices to which it belongs.  The exponent $\beta$ is an adjustable parameter within the simulations.  $S_{ER}$ is the Einstein-Regge action \cite{Regge:1961px} of discretized gravity,
 \begin{equation} \label{eq:GeneralEinstein-ReggeAction}
S_{ER}=-\kappa\sum_{j=1}^{N_2} V_{2}\delta_j+\lambda\sum_{j=1}^{N_4} V_{4},
\end{equation}
with $\kappa=(8\pi G)^{-1}$, $\lambda=\kappa\Lambda$, $\delta_j=2\pi-{\cal O}(t_j)\rm{arccos}(1/4)$ the deficit angle around a triangular hinge $t_j$, and with the volume of a $d$-simplex of equilateral edge length $a$ given by
\begin{equation} \label{eq:SimplexVolume}
V_{d}=\frac{\sqrt{d+1}}{d!\sqrt{2^{d}}}a^d.
\end{equation}
It is standard to absorb the overall numerical factors into constants and to perform the sums in Eq.~(\ref{eq:GeneralEinstein-ReggeAction}) so that the lattice action is given the convenient form
\bea\label{eq:ER}  S_{ER}=-\kappa_2 N_2+\kappa_4N_4,
\eea
with $N_4$ the number of four simplices and $N_2$ the number of triangles.

The lattice geometries are constructed by gluing together four-simplices along their ($4-1$)-dimensional faces.  The four-simplices are equilateral, with constant edge length $a$, and the dynamics is encoded in their connectivity.  In practice one would like to simulate for a fixed bare cosmological constant, but this is impractical since the simulations would take an exponentially long time to make excursions to large four-volumes.  Instead, it is standard in dynamical triangulations to simulate at fixed lattice volume, adding a volume preserving term to the action such as $\delta\lambda|N_4^f-N_4|$.  This term keeps the lattice volume in the simulations close to the target fiducial volume $N_4^f$.  In principle one should take the limit where $\delta \lambda$ is sent to zero, though in practice it suffices to take it sufficiently small.  Once the volume is fixed, the value of the parameter $\kappa_4$ is then completely fixed, as discussed in the following section.  The other parameters, $\kappa_2$ and $\beta$, form a two-dimensional parameter space for the phase diagram in which to search for a fixed point.

\begin{figure}
\begin{center}
\includegraphics[scale=.55]{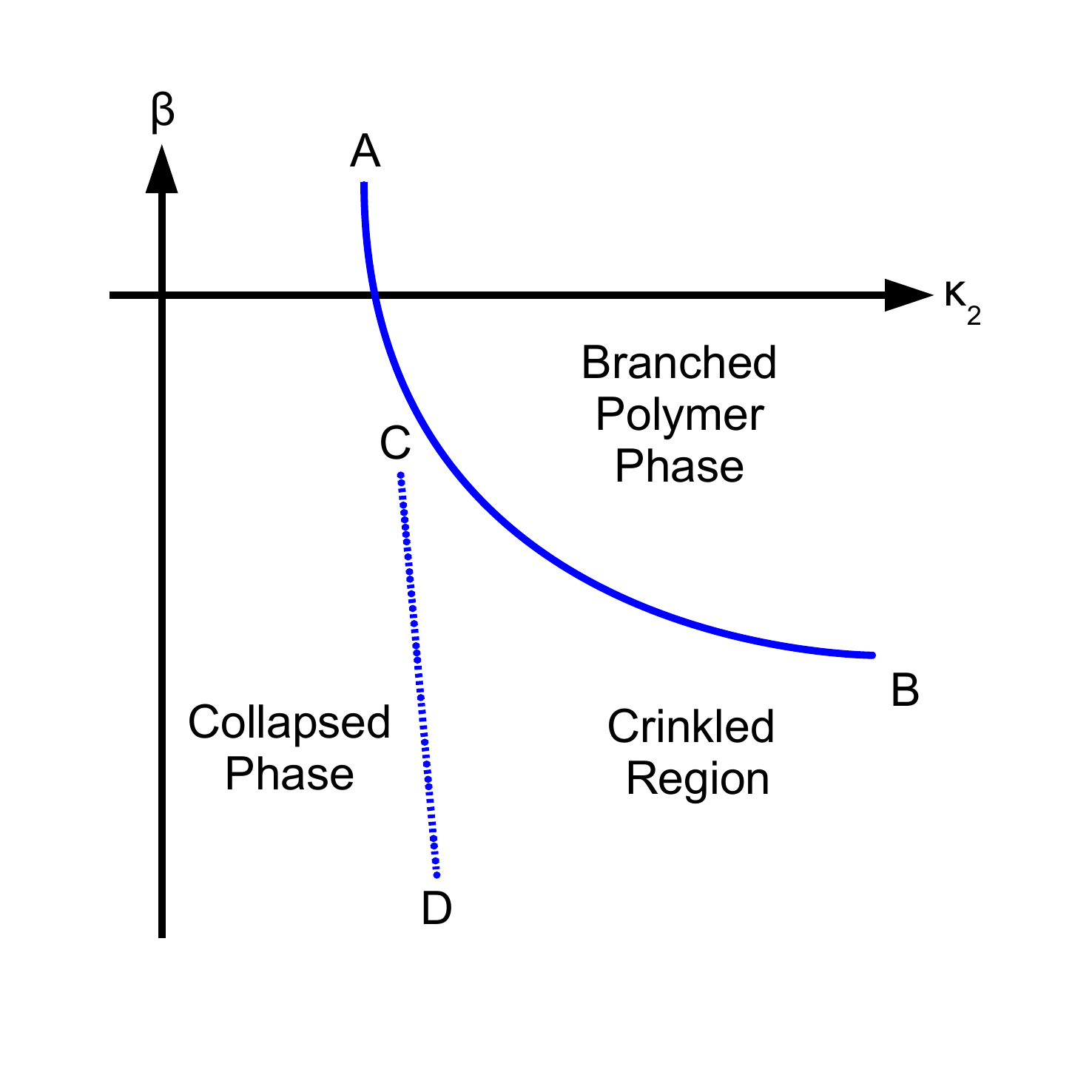}
\vspace{-3mm}
\caption{Schematic of the phase diagram as a function of $\kappa_2$ and $\beta$. \label{fig:phase1}}
\end{center}
\end{figure}

The phase diagram for this model has been mapped out in previous work \cite{Coumbe:2014nea}, and it is shown in Figure~\ref{fig:phase1}.  The solid line $AB$ is a first order transition line that separates the branched polymer phase from the collapsed phase.  Neither of these phases resembles semiclassical gravity.  The branched polymer phase has Hausdorff dimension 2, while the collapsed phase has a large, possibly infinite, dimension.  The crinkled region and the collapsed phase do not appear to be distinct phases; rather the crinkled region appears to be connected to the collapsed phase by an analytic cross-over.  The crinkled region requires very large volumes to see the characteristic behavior of the collapsed phase, suggesting that it is a part of the collapsed phase with especially large finite-size effects \cite{Coumbe:2014nea, Ambjorn:2013eha}.

It was shown in Ref.~\cite{Laiho:2016nlp} that a fine-tuning of $\beta$, such that one approaches the first-order transition line from the left, leads to semi-classical geometries with a dimension close to four.  The continuum limit appears to exist and is approached by following the transition line to large, possibly infinite, $\kappa_2$.

\section{The de Sitter instanton}
\label{sec:desitter}

The overall shape of our lattice geometries resembles that of Euclidean de Sitter space, and this agreement gets better as we take the continuum limit, as shown in Ref.~\cite{Laiho:2016nlp}.  Though this appears promising, we can do better by looking at the semiclassical approximation of the EDT partition function about the classical de Sitter solution.  The first part of the discussion in this section mirrors that of Ref.~\cite{Ambjorn:2012jv}, and following that we present our strategy for testing the expected behavior using our simulations.

Consider the partition function of Euclidean dynamical triangulations, where we assume that the sum over triangulations of fixed four-volume has already been performed, and we are only interested in the dependence on $N_4$.  Then the path integral defining the partition function reduces to a sum over $N_4$.  The leading behavior of the partition function is exponential in the four-volume
\bea \label{eq:Z}  Z(\kappa_4, \kappa_2)= \sum_{N_4} e^{-(\kappa_4 - \kappa_4^c)N_4}f(N_4, \kappa_2),
\eea
where $f(N_4, \kappa_2)$ is sub-exponential in $N_4$, and $\kappa_4^c$ is the pseudo-critical value of the coupling $\kappa_4$.  The limit $\kappa_4 \to\kappa_4^c$ allows one to take the infinite lattice-volume limit $N_4\to\infty$.  This is not necessarily the infinite physical-volume limit, since this procedure is equally valid in the unphysical crumpled phase, where the numerical simulations show that the emergent geometries are on the order of the size of the cut-off.  The critical value $\kappa_4^c$ is not known a priori, but emerges from the nonperturbative sum over triangulations.  In practice it is determined by adjusting the constant $\kappa_4$ at a particular target volume until the moves are equally likely to cause an upward fluctuation in volume as a downward one.  

This term in the exponential corresponds in the continuum to the renormalized cosmological constant term, so that we can identify
\bea \label{eq:cc} (\kappa_4-\kappa_4^c)N_4 = \frac{\Lambda}{8\pi G}V,
\eea
where $V=C_4 N_4 a^4$, with $C_4$ a geometric factor equal to $\sqrt{5}/96$.  Once the bare parameters $\kappa_2$ and $\beta$ are chosen such that the simulations are in the physical region of the phase diagram, the size of the semiclassical universe is specified when we input the target volume $N_4$.  The size of the de~Sitter universe at a given $\kappa_2$ and $\beta$ uniquely fixes $\kappa_4$, and thus the renormalized cosmological constant $\Lambda$.

 If the partition function in Eq.~(\ref{eq:Z}) is to reproduce semiclassical gravity, the subleading exponential behavior should be given by the Einstein-Hilbert term.  By power counting, the 4-volume dependence of this term should scale like 
 \bea \frac{1}{16\pi G}\int d^4x\sqrt{g}R \propto \frac{\sqrt{V}}{G}.
 \eea
Thus, the partition function with all other degrees of freedom integrated out except for the four-volume should have the form \cite{Ambjorn:2012jv}
\bea   Z(\kappa_4, \kappa_2)=\sum_{N_4} e^{-(\kappa_4 - \kappa_4^c)N_4 + k(\kappa_2)\sqrt{N_4}},
\eea
where the expected scaling of $k$ is
\bea \label{eq:k} k(\kappa_2) \propto \frac{a^2}{G}.
\eea
If a continuum limit exists, then for some values of the bare parameters $k$ approaches zero in lattice units, and $N_4$ approaches infinity, leaving the volume fixed in physical units.  The constant $k$ at a given lattice spacing must be determined from the simulations.  To see how, we consider the expectation value of the number of four-simplices, $\langle N_4 \rangle$, which a straightforward saddle-point expansion shows to be \cite{Ambjorn:2012jv}
\bea \label{eq:saddlepoint} \langle N_4 \rangle = \frac{\sum_{N_4}N_4 e^{-(\kappa_4 - \kappa_4^c)N_4 + k(\kappa_2)\sqrt{N_4}}}{\sum_{N_4}e^{-(\kappa_4 - \kappa_4^c)N_4 + k(\kappa_2)\sqrt{N_4}}} \nonumber \\  \approx \frac{k^2(\kappa_2)}{4(\kappa_4-\kappa_4^c)^2}.
\eea
In our simulations we fix $N_4$, so the expectation value $\langle N_4 \rangle=N_4$ is just an input to our simulations.  Solving Eq.~(\ref{eq:saddlepoint}) for $k$ we find
\bea \label{eq:slope}  k=2|\kappa_4-\kappa_4^c|\sqrt{N_4},
\eea
and from this we see that a plot of $\kappa_4$ as a function of $1/\sqrt{N_4}$ should be linear if the semi-classical limit is realized in the simulations.  Thus, a finite-volume scaling study of $\kappa_4$ should allow us to determine the slope of that linear dependence and thereby determine $k$.  Given $k$ we can get $G/a^2$, since by Eq.~(\ref{eq:k}), they are inversely proportional.  It remains to find that proportionality constant, which we get from the following argument.

The same saddle-point expansion used in Eq.~(\ref{eq:saddlepoint}) gives for the partition function
\bea \label{eq:HawkingMoss} Z(\kappa_4, \kappa_2) \approx \exp\left( \frac{k^2(\kappa_2)}{4(\kappa_4-\kappa_4^c)}\right) =\exp\left(\frac{3\pi}{G\Lambda}\right),
\eea
where the equality comes from a calculation of the continuum partition function, assuming that it is dominated by the de Sitter instanton.  This continuum expression for the partition function is the well-known Hawking-Moss instanton production amplitude \cite{Hawking:1981fz}.  In making this equality we assume that the approximate agreement between our lattice geometries and the continuum de Sitter solution gets better and better in the continuum limit, such that the de Sitter instanton dominates the partition function in that limit.  This picture can be tested, first by seeing whether $\kappa_4$ plotted versus $1/\sqrt{N_4}$ is linear, and second by computing the renormalized Newton's constant and comparing it to our recent determination of $G$ from the binding energy of scalar particles \cite{Dai:2021fqb}.  One can obtain the renormalized Newton's constant $G$ from the semiclassical partition function as follows.  Combining Eqs.~(\ref{eq:cc}), (\ref{eq:slope}), and (\ref{eq:HawkingMoss}), one finds
\bea G= \frac{5^{\frac{1}{4}}a^2}{16\sqrt{N_4}|\kappa_4-\kappa_4^c|},
\eea
which implies
\bea  \frac{G}{a^2}=\frac{5^{\frac{1}{4}}}{16|s|},
\label{eq:NCa}
\eea
with $s$ the slope determined by a fit to $\kappa_4$ as a function of $1/\sqrt{N_4}$.  

In practice we must calculate this slope for each new pair of values of $\kappa_2$ and $\beta$ in the bare action.  Even different volumes at the same nominal lattice spacing require separate additional volume runs to perform the finite-volume scaling, since the bare parameters vary to follow the transition line, which moves as a function of volume.  Each slope determined from finite-volume scaling at fixed values of $\kappa_2$ and $\beta$ determines a value of $G$ at some finite volume and nonzero lattice spacing.  These values of $G$ must then be extrapolated to the continuum, infinite volume limit.  

There is another subtlety involved in this analysis.  We want $G$ in the same physical units across ensembles, but our relative lattice spacing is given in simplex units $\ell$ rather than link units $a$, normalized at our fiducial lattice spacing ($\beta=0$).  The above prescription gives us $G$ in link units $a$, so if we want to put $G$ into common units across lattice spacings using the relative lattice spacing in simplex units $\ell$, then we must first convert $G$ into simplex units.  We also want to compare our final value for $G$ with that determined from the Newtonian binding of scalar particles given in Ref.~\cite{Dai:2021fqb}, but this determination involved scalar fields living on the simplices, and their correlation functions were measured on the dual lattice, making the corresponding masses, and thus $G$, in simplex units.  Therefore, we must use the conversion
\bea  \frac{G}{\ell_{\rm fid}^2}= \frac{G}{a^2}\left(\frac{a}{\ell}\right)^2\ell_{\rm{rel}}^2,
\label{eq:Gsimpl}
\eea
where the lattice spacing conversion factors are given in Table~\ref{tab:ensembles}, and $\ell_{\rm rel}=\ell/\ell_{\rm fid}$, with $\ell_{\rm fid}$ the fiducial lattice spacing in simplex units at $\beta=0$.  We discuss the determination of the lattice spacing conversion factors $a/\ell$ in Section~\ref{sec:num}.

\section{Details of the simulations}
\label{sec:sim}

The generation of the EDT ensembles is described in detail in Ref.~\cite{Laiho:2016nlp}, but we review some of those details here.  The lattices used in that and subsequent works \cite{Catterall:2018dns, Dai:2021fqb} have been saved and are reused here.  The path-integral sum is over a set of degenerate triangulations, where the usual combinatorial manifold constraints are relaxed \cite{Bilke:1998bn}.  Thus, distinct four-simplices may share the same five distinct vertex labels, and the neighbors of any given four-simplex are not necessarily unique.  Either of these conditions is a violation of the combinatorial manifold constraints.  However, degenerate triangulations have an advantage; there is a factor of $\sim$10 reduction in finite-size effects compared to combinatorial triangulations \cite{Bilke:1998bn}.  Since it is likely that degenerate and combinatorial triangulations are in the same universality class, if a continuum limit does in fact exist, we continue to use degenerate triangulations.

The numerical methods used to perform the simulations are by now well established \cite{Ambjorn:1997di}.  The standard (scalar) algorithm to perform the Monte Carlo integration of the path integral consists of an ergodic set of local moves, known as the Pachner moves, which are used to update the geometries \cite{Agishtein:1991cv, Gross:1991je,Catterall:1994sf}, and a Metropolis step, which is used to accept or reject the proposed move.  Most of the lattice ensembles used in this work were generated using a parallel variant of the standard algorithm, called parallel rejection.  This algorithm gives identical results, configuration by configuration, to the scalar algorithm, but the parallel streams can lead to a significant speed-up of the calculation.  Parallel rejection takes advantage of, and partially compensates for, the low acceptance of the Metropolis step in our simulations and is described in more detail in Ref.~\cite{Laiho:2016nlp}.  The sum over geometries is restricted to the fixed global topology $S^4$.  In order to enforce this restriction it is sufficient to start from the minimal four-sphere at the beginning of the Monte Carlo evolution, since the local moves are topology preserving.

Table~\ref{tab:ensembles} shows the ensembles that have been generated previously that are used in the present work.  They include ensembles at several different physical volumes and lattice spacings.  The relative lattice spacing quoted here was obtained in Ref.~\cite{Laiho:2016nlp} (and updated in Ref.~\cite{Dai:2021fqb}) by looking at the return probability of a diffusion process on the lattice geometries.  The return probability is dimensionless, but varies as a function of the diffusion time step, which is not.  One can rescale the diffusion time step so that the return probability lies on a universal curve; the rescaling factor then leads to the relative lattice spacing.  The errors quoted reflect the uncertainties in matching the curves in this procedure.  The ratio of link length $a$ and simplex distance $\ell$ on each ensemble is also given in Table~\ref{tab:ensembles}; the determination of this ratio is described in detail in the next section.
\begin{table}
\begin{center}
\begin{tabular}{cccccc}
\hline \hline
\ \ $\ell_{\rm rel}$ & $a/\ell$ \ \ & \ \ $\beta$ \ \ \ & \ $\kappa_2$ \ & \ \ \ \ \ $N_4$ \ \ & \ \ \ Number of configs \\
\hline
1.59(10) & 3.4(3)  & 1.5 & 0.5886 & \ \ 4000 & 367  \\
1.28(9) & 3.9(2) & 0.8 & 1.032 & \ \ 4000 &  524  \\
1 &  5.2(1) & 0 & 1.605 &         \ \ 2000 & 248  \\
1 &  5.2(1) & 0 & 1.669 &         \ \ 4000 & 575  \\
1 &  5.2(1) & 0 & 1.7024 &     \ \  8000 & 489  \\
1 &  5.2(1) & 0 & 1.7325 &     \ \ 16000 & 501  \\
1 &  5.2(1) & 0 & 1.75665 &     \ \ 32000 & 1218  \\
0.80(4) & 7.2(7) & $-0.6$ & 2.45 & \ \ 4000 & 414  \\
0.70(4) & 8.6(9) & $-0.8$ & 3.0 &  \ \ 8000 & 1486  \\
0.70(4) & 8.6(9) & $-0.776$ & 3.0 & \ \ 16000 & 2341 \\
\hline
\end{tabular}
\caption{The parameters of the ensembles used in our studies of EDT.  The first column shows the relative lattice spacing in units of simplex distance, with the ensembles at $\beta=0$ serving as the fiducial lattice spacing.  The quoted error is a systematic error associated with matching the return probabilities across lattice spacings.  The second column shows the ratio of the link distance $a$ to the simplex distance $\ell$ on a given ensemble, with a systematic error associated with the matching procedure.  All $a/\ell$ have been corrected for finite size effects.  The third column is the value of $\beta$, the fourth is the value of $\kappa_2$, the fifth is the number of four-simplices in the simulation, and the sixth is the number of configurations sampled.}
\label{tab:ensembles}
\end{center}
\end{table}

Table~\ref{tab:ensemblesNew} shows the new ensembles created in this work to perform the finite-size scaling of $\kappa_4$ for fixed values of the other two parameters $\kappa_2$ and $\beta$.  As the volume of the lattice is increased at fixed lattice spacing, one of these other two parameters must be re-tuned in order to move the simulation closer to the phase transition line, which shifts as the volume changes, even for what is nominally the same lattice spacing.  Thus, we need a series of additional runs at different volumes even where we already have multiple volumes at the same nominal lattice spacing.  Since the phase transition line shifts to the right for increasing volume, and to the left for decreasing volume, it is necessary to go to larger volumes than the nominal volume when we do the finite-size scaling, since only in that case do we remain in the correct phase.  Thus, all of the volumes used in the finite-size scaling of $\kappa_4$ are larger than the nominal volume of our original ensembles.
\begin{table}
\begin{center}
\begin{tabular}{ccccc}
$\beta$  & $\kappa_2$ &  $N_4$  &  $\#$ of configs & $\kappa_4$\\
\hline\hline\\[-8pt]
1.5 & 0.5886  &  4000  & 414  & 7.989973(93)\\
   &       &  8000  & 327  & 7.99258(18)\\
   &       &  16000 & 801  & 7.99530(27)\\
   &       &  32000 & 584  & 7.996832(49)\\
   &       &  64000 & 494  & 7.997903(77)\\
\hline\hline\\[-8pt]
0.8 & 1.032  &  4000  & 262  & 7.00464(11)\\
  &   &  8000  & 495  & 7.00800(18)\\
    &      &  16000 &  91  & 7.01003(15)\\
     &     &  32000 & 369  & 7.011645(77)\\
      &    &  64000 & 869  & 7.012781(43)\\
\hline\hline\\[-8pt]
0   & 1.605      &   2000 & 1712  & 6.147791(67)\\
   &       &   4000 &  414  & 6.152958(79)\\
 &          &   6000 &  579  & 6.154980(81)\\
 &         &   8000 &  327  & 6.15600(12)\\
  &        &  12000 &  244  & 6.15733(12)\\
   &       &  16000 &   28  & 6.15800(31)\\
\hline\hline\\[-8pt]
 0  & 1.669      &   4000 &    476  & 6.32841(18)\\
      &    &   8000 &   2849  & 6.330489(58)\\
      &    &  16000 &   1216  & 6.332214(59)\\
      &    &  32000 &   1208  & 6.333493(49)\\
      &    &  64000 &   903  & 6.33420(11)\\
\hline\hline\\[-8pt]
0     & 1.7024    &   8000 &   489  & 6.42259(18)\\
    &      &  12000 &   1056  & 6.424000(58)\\
    &      &  16000 &   1145  & 6.424592(59)\\
    &      &  32000 &   1529  & 6.425800(49)\\
    &      &  64000 &   295  & 6.42652(11)\\
\hline\hline\\[-8pt]
0    & 1.7325     &  16000 &   402  & 6.50854(20)\\
      &    &  24000 &   430  & 6.509460(96)\\
      &    &  32000 &   1369  & 6.509929(58)\\
      &    &  64000 &   95  & 6.510592(73)\\
\hline\hline
-0.6    & 2.45     &  4000 &  414   & 6.78342(25) \\
      &    &  8000 &  298   &  6.78545(15) \\
      &    &  12000 &  807   &  6.786175(94)\\
      &    &  16000 &   973  &  6.786636(98)\\
       &    &  24000 &    1057 &  6.787203(78)\\
       &    &  32000 &    343 &  6.78758(15)\\
\hline\hline
\end{tabular}
\caption{The parameters of the ensembles used to extract the volume scaling of $\kappa_4$.  The first three columns label the ensembles.  The first column is $\beta$, the second is $\kappa_2$, and the third is the lattice volume $N_4$.  The fourth column is the number of configurations used to determine $\kappa_4$, and the fifth column is the value of $\kappa_4$ determined on that ensemble, along with its statistical error. }
\label{tab:ensemblesNew}
\end{center}
\end{table}

\section{Numerical results}
\label{sec:num}

\subsection{Relating lattice distance measurements}

We present the calculation of the conversion factors between link units $a$ and simplex units $\ell$ on our EDT ensembles.  First, we review the calculation of the return probability $P(\sigma)$, with diffusion time $\sigma$, on the dual lattice.  This quantity has been used to set the relative lattice spacing in previous works \cite{Laiho:2016nlp, Dai:2021fqb}.  Before starting the random walk of the diffusion process on the dual lattice, the lattice is first shelled, with a starting four-simplex chosen at random as the source; the next shell consists of the nearest neighbors of the source simplex.  The next shell consists of all of their nearest neighbors, without replacement, and so on until all of the four-simplices of the lattice configuration have been counted.  The starting simplex for the diffusion process is then chosen from the shell with the maximum number of four-simplices.  We find that restricting our sources to come from the largest three-slice minimizes finite lattice spacing effects, and it is the same procedure that has been used throughout the recent EDT work involving the present authors, including the study of K\"{a}hler-Dirac fermions \cite{Catterall:2018dns} and the study of scalar interactions \cite{Dai:2021fqb}.

The diffusion process on the dual lattice uses a random walk where the next jump is chosen from the neighbors of a given simplex.  Because degenerate triangulations are used, some of the five neighbors of a four-simplex are not unique, that is, sometimes the same four-simplex shares multiple tetrahedra with a neighboring four-simplex.  Even so, each of the five neighbors of a given four-simplex is given equal weight when choosing the next step of the random walk.  One source is used per configuration, and many random walks starting from that source are run in order to sample the probability of returning to the starting four-simplex.  One peculiarity of degenerate triangulations is that for the dual lattice return probability, all of the odd time steps have zero probability, at least for time steps sufficiently early in the diffusion process.  In order to compute the return probability, and the corresponding spectral dimension, we take only the even time steps, so that each step $\sigma$ is actually two lattice hops in the diffusion process.  This procedure of omitting the odd steps in the return probability was shown to work in the branched polymer phase, where it correctly reproduces the known spectral dimension of $4/3$ \cite{Coumbe:2014nea}.  This procedure was also used to compute the return probability and spectral dimension in the subsequent work on the tuned semi-classical geometries \cite{Laiho:2016nlp}.

In order to get the ratio of the link distance and the simplex distance, we compare the return probability on the direct lattice with that on the dual lattice.  The implementation of the diffusion process on the direct lattice is new to the present work.  Since the hops are now between vertices, and each vertex is separated by link length $a$, this allows us to convert simplex distance to link distance.  The random walk used to compute the return probability is once again chosen from the shell with the maximal volume, but this time the shelling is performed on the vertices.  In the diffusion process, a given vertex does not have a fixed number of neighbors.  In fact, the number of neighbors can occasionally grow to be quite large.  For this reason it is helpful to use dynamical memory allocation while computing the diffusion process.  For this work, an array of linked lists was used to store all of the neighboring vertices to any particular vertex on a given configuration.  Because the triangulations are degenerate, there can exist multiple links connecting the same two vertices.  All such links are given equal weight when computing the probability of a hop to a nearest neighbor.  In the case of the return probability on the direct lattice, both even and odd diffusion time steps are non-zero and are used in the calculation.  There is an oscillation visible between the even and odd steps at early times due to discretization effects; this oscillation dies out after a sufficiently large number of time steps.  This effect is common in computations involving the return probability or spectral dimension on random lattices \cite{Ambjorn:2005qt}.

\begin{figure}
    \centering
    \includegraphics[origin=c, width=12cm]{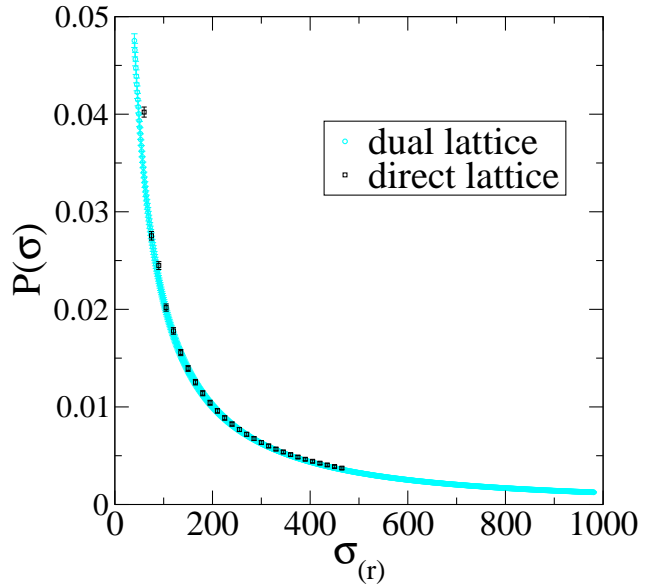}
    \caption{The return probability $P(\sigma)$ as a function of the diffusion step size $\sigma$ for both the dual lattice and the direct lattice at a volume of $N_4=32,000$ and $\beta=0$.  The return probability for the direct lattice has a rescaled $\sigma_r$ so that it overlaps with the return probability of the dual lattice. }
    \label{fig:a_rel_32}
\end{figure}

\begin{figure}
    \centering
    \includegraphics[origin=c, width=12cm]{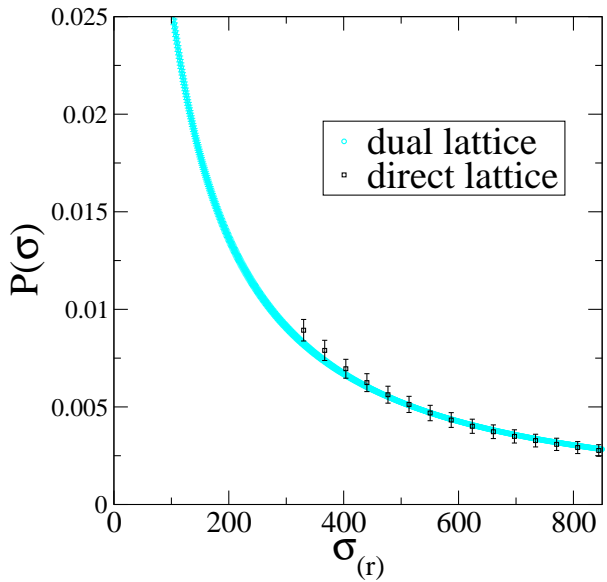}
    \caption{The return probability $P(\sigma)$ as a function of the diffusion step size $\sigma$ for both the dual lattice and the direct lattice at a volume of $N_4=4000$ and $\beta=-0.6$.  The return probability for the direct lattice has a rescaled $\sigma_r$ so that it overlaps with the return probability of the dual lattice.}
    \label{fig:a_rel_4}
\end{figure}

Figure~\ref{fig:a_rel_32} shows the return probabilities for both the dual and direct lattice diffusion processes on the 32k, $\beta=0$ ensemble.  The return probability on the direct lattice has been rescaled along the $\sigma$ axis so that the two curves overlap.  This rescaling factor is used to determine the ratio $a/\ell$.  Recalling that the diffusion step is proportional to distance squared, calling $\sigma_{\rm dual}$ the diffusion time step on the dual lattice, and $\sigma_{\rm direct}$ the diffusion time step on the direct lattice, we find 
\bea  \frac{a}{\ell} = \sqrt{\frac{2\sigma_{\rm dual}}{\sigma_{\rm direct}}}
\eea
where it is assumed that the $\sigma$s are at matching points on the return probability curve.  The factor of 2 accounts for the fact that each step of the diffusion process on the dual lattice is actually two lattice hops.  As can be seen in Fig.~\ref{fig:a_rel_32}, the agreement between the rescaled curves is very good.  Fig.~\ref{fig:a_rel_4} shows this same matching on the finer ensemble at $\beta=-0.6$.  Again, the rescaled curves line up nicely.  

\begin{figure}
    \centering
    \includegraphics[width=\linewidth]{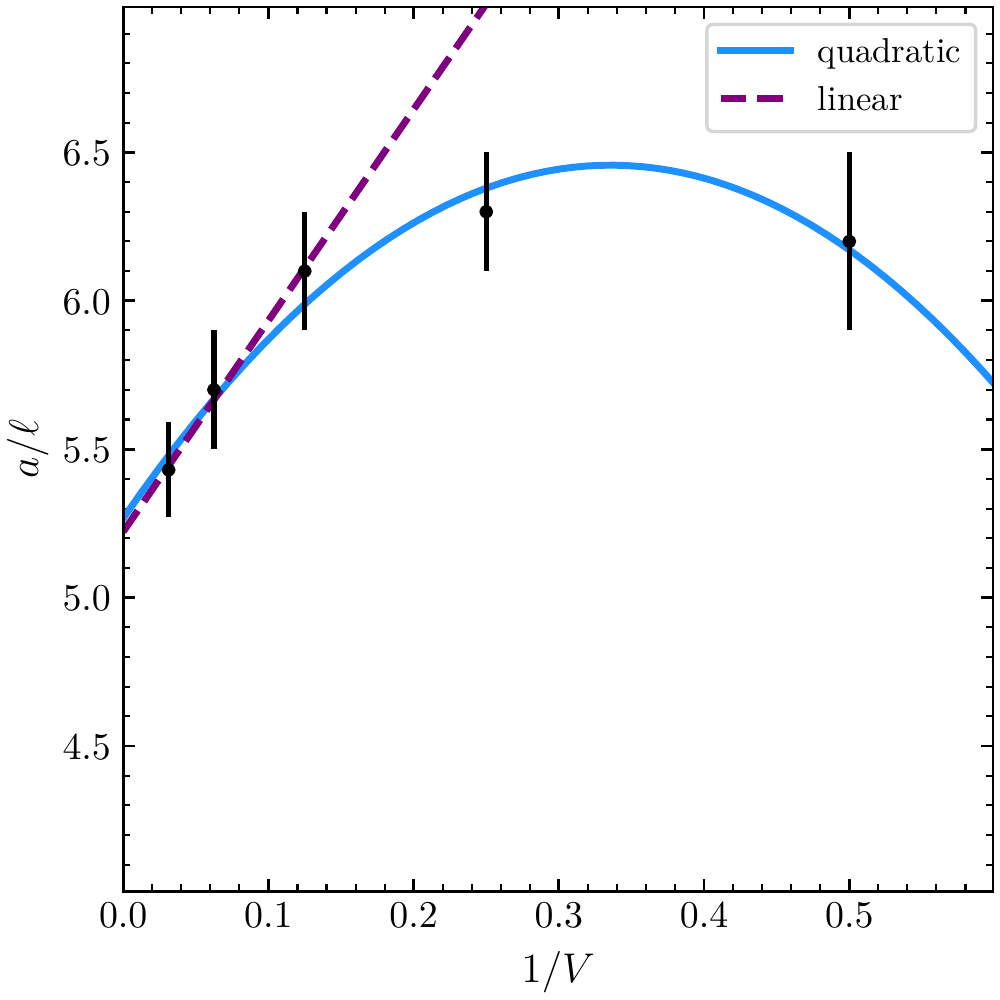}
    \caption{The ratio of direct to dual lattice spacings $a/\ell$ as a function of $1/V$ at $\beta=0$ for multiple volumes, and two sample fits extrapolating this quantity to the infinite volume limit.}
    \label{fig:a_rel_FV}
\end{figure}

Table~\ref{tab:ensembles_aoverl} presents our values for $a/\ell$ extracted from each of our ensembles.  For our Newton's constant analysis we quote a single number for $a/\ell$ at a given lattice spacing.  These values are corrected for finite-volume effects.  In the case of the $\beta=0$ ensembles, where we have multiple lattice volumes, we do a direct extrapolation to infinite volume.  This extrapolation is shown in Fig.~\ref{fig:a_rel_FV}.   In order to correct all of the values of $a/\ell$ at other lattice spacings for finite-volume effects, we assume that the finite volume dependence is the same as that of the $\beta=0$ ensembles, and we use that dependence to determine a correction factor for $a/\ell$.  This is done by matching the physical volume of the ensembles at other lattice spacings against those at $\beta=0$, and computing the percentage difference between where that physical volume lines up with the curve in Fig.~\ref{fig:a_rel_FV} and the infinite volume limit.

\begin{table}
\begin{center}
\begin{tabular}{ccc}
\hline \hline
\ \ $\ell_{\rm rel}$ & \ \ \ \ \ $N_4$ \ \ & \ \ \ $a/\ell$ \\
\hline
1.59(10) &  \ \ 4000 & \ \ \ 3.6(3) \\
1.28(9) &  \ \ 4000 & \ \ \ 4.3(2)  \\
1 &         \ \ 2000 & \ \ \ 6.2(3)  \\
1 &       \ \ 4000 & \ \ \ 6.3(2)  \\
1 &     \ \  8000 & \ \ \ 6.1(2)  \\
1 &      \ \ 16000 & \ \ \ 5.7(2)  \\
1 &      \ \ 32000 & \ \ \ 5.43(16)  \\
0.80(4)  & \ \ 4000 & \ \ \ 8.6(2)  \\
0.70(4)  &  \ \ 8000 & \ \ \ 10.6(6)  \\
0.70(4) &  \ \ 16000 & \ \ \ 10.4(5)  \\
\hline
\end{tabular}
\caption{The values of $a/\ell$ for the different ensembles in our analysis.  The first two columns identify the ensemble, the first by its relative lattice spacing in units of simplex distance, with the ensembles at $\beta=0$ serving as the fiducial lattice spacing.  The second column identifies the ensemble by the lattice volume.  The third column is the value of $a/\ell$ on that ensemble, with an error associated with matching the return probability curves.}
\label{tab:ensembles_aoverl}
\end{center}
\end{table}

\begin{table}
\begin{center}
\begin{tabular}{cc}
\hline \hline
\ \ $\ell_{\rm rel}$  & \ \ \ $a/\ell$ \\
\hline
1.59(10) &  \ \ \ 3.4(3) \\
1.28(9)  & \ \ \ 3.9(2)  \\
1 &  \ \ \ 5.2(1)  \\
0.80(4)  & \ \ \ 7.2(7)  \\
0.70(4)  & \ \ \ 8.6(9)  \\
\hline
\end{tabular}
\caption{The values of $a/\ell$ for different lattice spacings.  The first column identifies the ensemble by its relative lattice spacing in units of simplex distance.  The second column is the value of $a/\ell$ at that lattice spacing in the infinite volume limit, including the total error.}
\label{tab:aInfiniteVol}
\end{center}
\end{table}

The errors in the values of $a/\ell$ are estimated as follows.  First, the statistical errors are taken into account by varying the matching factor according to the $1\sigma$ statistical errors in the data points for the return probabilities.  Second, we account for the errors associated with extrapolating $a/\ell$ at a given lattice spacing to its value in the infinite volume limit.  At $\beta=0$, where the extrapolation to infinite volume can be done explicitly, we vary the fit form and the number of data points included in the fit in order to estimate a systematic error associated with the infinite-volume extrapolation.  Figure~\ref{fig:a_rel_FV} shows a quadratic fit to all five volumes at $\beta=0$ and a linear fit to the largest three volumes.  We also consider a quadratic fit to the four largest volumes.  Based on the spread in these results, we quote an infinite volume result of $a/\ell=5.2(1)$ at $\beta=0$.  The errors in the infinite-volume results for $a/\ell$ at other lattice spacings are obtained by combining the error in the finite volume correction with the error in $a/\ell$ at a given lattice spacing; the central values with their errors are quoted in Tab.~\ref{tab:aInfiniteVol}.

\subsection{Finite volume scaling and Newton's constant}
\begin{figure}
    \centering
    \includegraphics[width=\linewidth]{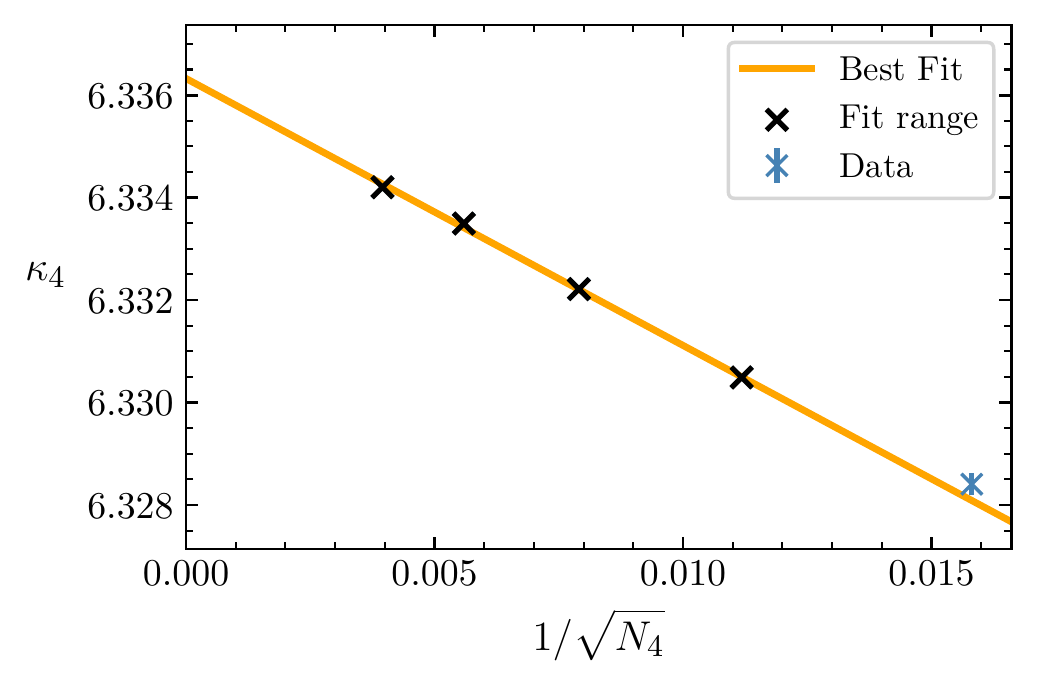}
    \caption{The $\kappa_4$ values corresponding to the parameters $\beta=0$ and $\kappa_2=1.669$ versus $1/\sqrt{N_4}$. The line is a linear fit with $\chi^2/\mathrm{d.o.f.}=1.4$ and $p$-value of $0.24$, resulting in a slope $s=0.521(11)$, cf.~Tab.~\ref{tab:ensemblesslopes}. The data point corresponding to the tuned ensemble, i.e., the smallest volume, was not included in the fit. }
    \label{fig:Beta04kextrapol}
\end{figure}

\begin{figure}
    \centering
    \includegraphics[width=\linewidth]{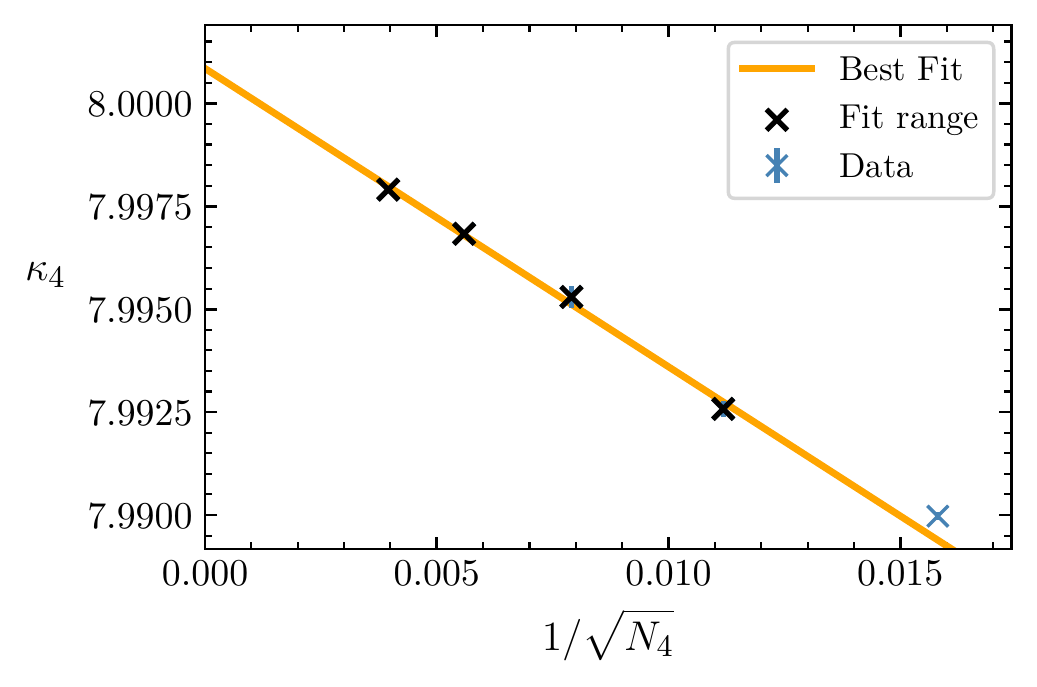}
    \caption{The $\kappa_4$ values at the coarsest lattice spacing with $\beta=1.5$ and $\kappa_2=0.5886$ versus $1/\sqrt{N_4}$. The line is a linear fit with $\chi^2/\mathrm{d.o.f.}=1.4$ and $p$-value of $0.24$, resulting in a slope of $s=0.724(32)$. The data point at the smallest volume was not included in the fit. }
    \label{fig:Beta1.5extrapol}
\end{figure}

\begin{figure}
    \centering
    \includegraphics[width=\linewidth]{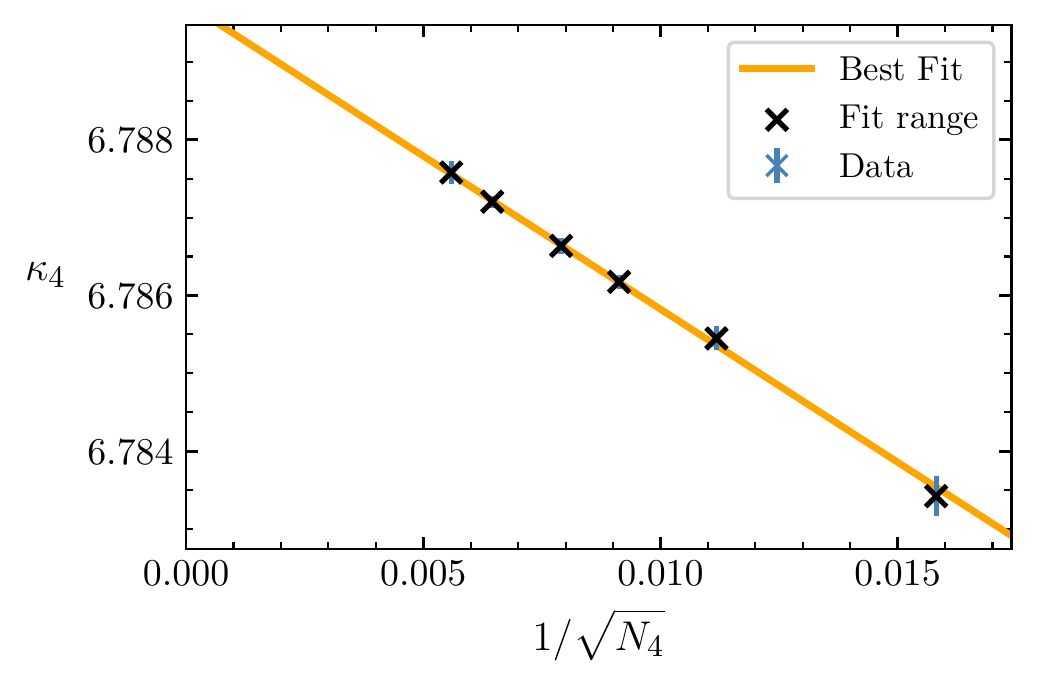}
    \caption{The $\kappa_4$ values corresponding to the parameters $\beta=-0.6$ and $\kappa_2=2.245$ versus $1/\sqrt{N_4}$. The line shows a linear fit with $\chi^2/\text{d.o.f.}=0.15$ and a $p$-value of 0.96, resulting in a slope $s=0.393(22)$.}
    \label{fig:Beta-0.6extrapol}
\end{figure}

To compute Newton's constant given in Eq.~(\ref{eq:NCa}) in link units $a$, we need to extract the value of ${\sqrt{N_4}|\kappa_4-\kappa_4^{c}|}$, with the pseudo-critical value $\kappa_4^{c}$. The parameter $\kappa_4^{c}$ corresponds to the value of the coupling $\kappa_4$ needed to take the infinite lattice-volume limit, and it is a function of $\beta$ and $\kappa_2$. To extract $\sqrt{N_4}|\kappa_4-\kappa_4^{c}|$, we therefore perform simulations at fixed values of $\kappa_2$ and $\beta$, and increasing volumes $N_4$, and measure the tuned value of $\kappa_4$ at each of the volumes. A linear fit 
\begin{equation}
\label{eq:kappa4fit}
    \kappa_4(N_4)=A_{\kappa_4}+s\,\frac{1}{\sqrt{N_4}}\,,
\end{equation}
then allows us to determine the slope $s$ to obtain Newton's constant $G/a^2$ in link units. The errors on the values for $\kappa_4(N_4)$ are assumed to be purely statistical, and are estimated using single-elimination jackknife resampling. Autocorrelation errors are taken into account by a blocking procedure, where $\kappa_4$ data sets are blocked until the error stops increasing. The $\kappa_4$ values of all ensembles are summarized in Tab.~\ref{tab:ensemblesNew}, where the smallest volume at fixed values of $\beta$ and $\kappa_2$ corresponds to the tuned ensemble close to the first order phase transition on which measurements of physical quantities have been performed in previous works. Tab.~\ref{tab:ensemblesslopes} summarizes the resulting slopes $|s|$ for each pair of $\kappa_2$ and $\beta$ values, together with the $\chi^2/\mathrm{d.o.f.}$ and the $p$-value corresponding to each fit. In Figs.~\ref{fig:Beta04kextrapol}, \ref{fig:Beta1.5extrapol}, and \ref{fig:Beta-0.6extrapol}, we display examples of these fits, each of them showing a different finite-volume scaling study at a different relative lattice spacing $\ell_{\text{rel}}$. In some of these fits, the $\kappa_4$ value of the tuned ensemble, i.e., the smallest volume for each set of values for $\kappa_2$ and $\beta$, was discarded due to it not being well described by a linear fit to the rest of the data points.  A possible reason for this is that the lattice volume is too small, or that its  closeness to the first-order phase transition results in contamination by occasional tunneling into the branched polymer phase, where the values of $\kappa_4$ differ significantly \cite{Coumbe:2014nea}.  We find good evidence of linear scaling of $\kappa_4$ as a function of $1/\sqrt{N_4}$ across lattice spacings and nominal volumes, showing strong numerical evidence for the validity of the semi-classical approximation, Eq.~(\ref{eq:slope}). 

\begin{table}
\begin{center}
\begin{tabular}{ccccccc}
\hline \hline
\ \ $\ell_{\mathrm{rel}}$\ \ & \ \ $V_{\rm rel}$ \ \ &\ \ $\beta$ \ \ & \ \ $\kappa_2$ \ \ & \ \ $|s|$\ \ &\ \ $\chi^2/\mathrm{d.o.f.}$\ \ & p-value \\
\hline\\[-8pt]
1.59(10)& 25.6(6.4) &1.5 &0.5886& 0.724(32)&1.4 &0.24 \\
1.28(9)&10.7(3.0) &0.8& 1.032 & 0.6840(55)&0.35 &0.79 \\
1 & 2.0(0)& 0&1.605& 0.652(14)&0.60&0.62 \\
1 & 4.0(0)& 0&1.669& 0.521(11)&1.4 &0.24\\
1 & 8.0(0)& 0&1.7024& 0.502(12)&0.43 &0.65\\
1 & 16.0(0)& 0&1.7325& 0.436(39)&0.76 &0.38\\
0.80(4) &1.64(32)& -0.6&2.45&0.393(22) &0.15 &0.96\\
\hline
\end{tabular}
\caption{Extracted slopes $s$ following a fit of the data in Tab.~\ref{tab:ensemblesNew} to Eq.~(\ref{eq:kappa4fit}), together with the other relevant parameters of the ensembles.  The first column is the relative lattice spacing in simplex units. The second is the relative physical volume, given by $V_{\rm rel}=N_4 \ell_{\text{rel}}^4$, in units of thousands of four-simplices.  The third and fourth columns are $\beta$ and $\kappa_2$, respectively.  The fifth column is the slope $|s|$, the sixth column is the $\chi^2/{\rm d.o.f.}$ of the linear fit, and the seventh column is the p-value of that fit. }
\label{tab:ensemblesslopes}
\end{center}
\end{table}

Given the finite-volume scaling of $\kappa_4$, the conversion factor for link units into simplex units $a/\ell$, and the relative lattice spacing in simplex units $\ell_{\rm rel}$, we compute a value of Newton's constant in units of our fiducial lattice spacing using Eq.~(\ref{eq:Gsimpl}) for all of our results across lattice spacings and nominal volumes. To perform the extrapolation to the continuum, infinite volume limit, we use the simplest viable ansatz for the dependence of $G$ on the physical volume and lattice spacing. We use a fit function for the Newton constant extrapolation that is similar to  what was used in the recent study of gravitational binding \cite{Dai:2021fqb},
\begin{align}
\label{eq:G-fit}
    G = \frac{H_{G}}{V} + I_{G} \ell_{\text{rel}}^{2} + J_{G} \ell_{\text{rel}}^{4} + K_{G},
\end{align}
where $H_{G}$, $I_{G}$, $J_{G}$, and $K_{G}$ are fit parameters. We include quartic corrections in the relative lattice spacing, since the coarse lattices introduce curvature as a function of the relative lattice spacing.  The inclusion of $1/V^2$ corrections is not necessary, however, since the additional fit parameter does not improve the quality of the fit.  To test the results of the fit ansatz Eq.~(\ref{eq:G-fit}), we perform an additional fit with $J_G$ set to zero, while simultaneously dropping the data points with $\ell_{\mathrm{rel}}>1$. The result for the continuum, infinite-volume limit of $G$ for this fit is consistent within one sigma with that of the extrapolation using the full ansatz given in Eq.~(\ref{eq:G-fit}).

\begin{figure}
    \centering
    \includegraphics[width=\linewidth]{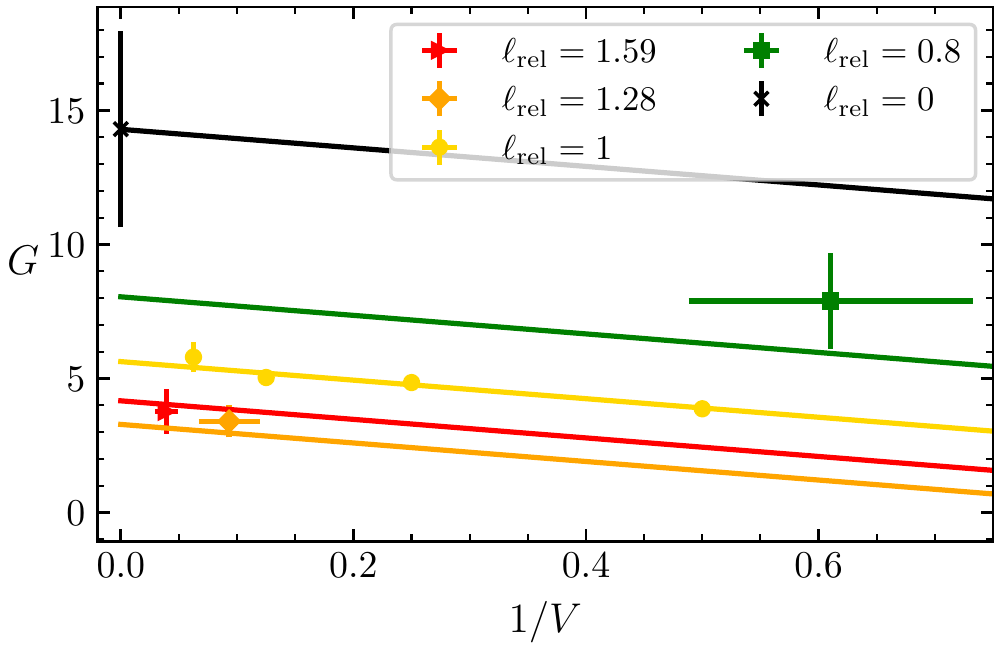}
    \caption{Newton's constant $G$ as a function of the inverse physical volume (expressed in units of 1000 simplices) for all of the ensembles (colored), as well as the continuum limit (in black).  Here quadratic corrections in $1/V$ as well as $\ell_{\text{rel}}^{2}$ were used to  model the extrapolation.  For this fit we find $\chi^{2}/\text{d.o.f.} = 0.87$ corresponding to a $p$-value of 0.46, and the continuum, infinite volume value is $G = 14.3(3.6)$.}
    \label{fig:GVolume}
\end{figure}

\begin{figure}
    \centering
    \includegraphics[width=\linewidth]{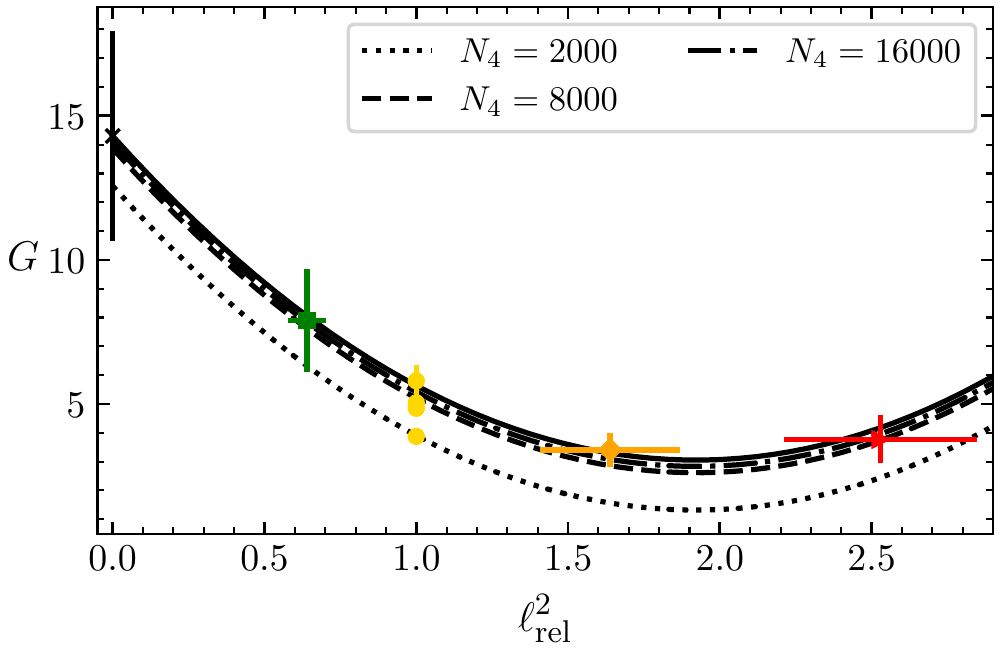}
    \caption{The same data and fit from Fig.~\ref{fig:GVolume} however now plotted as a function of the squared lattice spacing.  Here example lines of constant physical volume are plotted along with the infinite volume limit as a solid black line, and the data are represented in the same manner as Fig.~\ref{fig:GVolume}.}
    \label{fig:Gspacing}
\end{figure}

The extrapolation of $G$ is shown in Fig.~\ref{fig:GVolume} against the inverse physical volume. The colored lines correspond to lines of constant lattice spacing, and the black line represents the continuum limit extrapolation. Fig.~\ref{fig:Gspacing} shows the same data, plotted against the squared relative lattice spacing, which represents a different slice through the parameter space spanned by $1/V$ and $\ell_{\mathrm{rel}}$. Here, lines of fixed physical volume at the fiducial lattice spacing, i.e. $\ell_{\mathrm{rel}}=1$ are shown. In both figures, the black cross corresponds to the infinite-volume continuum extrapolation of the Newton constant.  We find for the extrapolated value of the Newton constant $G=14.3(3.6)$.  The $\chi^2/\mathrm{d.o.f.}$ for the extrapolation fit is $0.87$, corresponding to a $p$-value of $0.46$, an acceptable confidence level for this fit.  This result for $G$ in fiducial simplex units can be compared directly to our recent result from Newtonian binding in Ref.~\cite{Dai:2021fqb}, where the value $G=15(5)$ was quoted.  The agreement is clearly excellent, and it is a powerful check that both calculations are making contact with the correct semi-classical limit.

\section{Conclusion}
\label{sec:conclusion}

In this work we have revisited the emergence of de Sitter space in the EDT formulation.  We have studied whether the lattice geometries that emerge from our simulations are compatible with the semiclassical de Sitter solution in the continuum, large-volume limit.  Following the discussion in Ref.~\cite{Ambjorn:2012jv}, we have studied the saddle-point approximation of the Euclidean partition function about de Sitter space.
The finite-volume scaling of the bare cosmological constant in the semi-classical approximation can be shown to be a linear function of $1/\sqrt{V}$, where $V$ is the lattice volume.  Our data confirms this expectation.  Given this agreement, it is possible to use this result to extract a value of the renormalized Newton constant $G$ from a comparison between the lattice partition function in the semiclassical limit and a similar calculation in the continuum.  The continuum calculation in this case is the well-known Hawking-Moss instanton solution \cite{Hawking:1981fz}, evaluated for the special case of tunneling to a de Sitter universe.

This identification provides a value of $G$ at a series of volumes and lattice spacings, so that it is necessary to extrapolate these values to the continuum, infinite-volume limit.  Before doing so, we must put $G$ at the different lattice spacings into common physical units.  There is a subtlety here, in that we obtain our raw values of $G$ in link units, while our relative lattice spacings are determined in simplex units.  We calculate the conversion factor by comparing the return probabilities computed on the dual lattice and on the direct lattice.  With the appropriate conversion factors, and after the extrapolation, we finally find a value of $G=14.3(3.6)$, measured in simplex units at our fiducial lattice spacing at $\beta=0$.  This result can be compared to our previous calculation of Newton's constant obtained by studying the gravitational interaction of scalar particles.  Both calculations use the same tuned ensembles described in this work, and the value $G=15(5)$ given in Ref.~\cite{Dai:2021fqb} is normalized in the same units as the one presented here, so that a comparison is possible.  The agreement is clearly very good.  Our new result for $G$ implies a somewhat improved determination of our absolute lattice spacing in Planck units, with $\ell_{Pl}=(3.8 \pm 0.5) \ell_{\rm fid}$.

The main source of error in the determination of Newton's constant in the present analysis is the error in the conversion factor $a/\ell$ between link and simplex units, and the determination of the relative lattice spacing. The latter also determines the uncertainties on the physical volume and the squared lattice spacing, cf.~Figs.~\ref{fig:GVolume} and \ref{fig:Gspacing}. A reduction of the uncertainties on the quantities $a/\ell$ and $\ell_{\mathrm{rel}}$ in the future will most likely require larger volumes at finer lattice spacings.  Additional measurements of the finite-volume scaling of the bare cosmological constant at finer lattice spacings and larger volumes should also allow for a better determination of $G$ in the continuum, infinite-volume limit.  Improved precision on this quantity is important for testing the consistency of the EDT formulation.

In conclusion, the good agreement between the determination of Newton's constant in Ref.~\cite{Dai:2021fqb} and the present work is highly non-trivial, since one calculation studies the semi-classical expansion of the partition function about de Sitter space, and the other measures the gravitational interaction between scalar particles.  That both of these features emerge from Euclidean dynamical triangulations governed by the same universal constant provides strong evidence that EDT is not merely a theory of random geometry, but a theory of gravity.

\bigskip

\begin{acknowledgments}
JUY was supported by the U.S. Department of Energy grant DE-SC0019139, and by Fermi Research Alliance, LLC under Contract No. DE-AC02-07CH11359 with the U.S. Department of Energy, Office of Science, Office of High Energy Physics. SB and JL were supported by the U.S. Department of Energy (DOE), Office of Science, Office of High Energy
Physics under Award Number DE-SC0009998. MS was supported by the German Academic Scholarship Foundation and gratefully acknowledges hospitality at Syracuse University and at CP3-Origins, University of Southern Denmark, during different stages of this project.   Computations for this work were performed in part on facilities of the USQCD Collaboration, which are funded by the Office of Science of the U.S. Department of Energy. Computations were also carried out in part on the Syracuse University HTC Campus Grid and were supported by NSF award ACI-1341006.
\end{acknowledgments}

\bibliography{refs}

\end{document}